\documentclass[12pt,review,authoryear]{elsarticle}

\usepackage{gensymb} %for degree symbol

\usepackage{multirow}
\usepackage{tabularx}
\usepackage{longtable} %long tables
\usepackage{pdflscape}
\usepackage{url} %allows url in references to occupy more than 1 line

 \usepackage{xcolor}
\usepackage{soul}
\usepackage{lineno}

\usepackage[margin=3cm]{geometry}

\journal{Journal of Experimental Biology}

\begin{document}

\begin{frontmatter}


\title{Tibial Strains are Sensitive to Speed, but not Grade, Perturbations During Running}

\author[inst1,inst2]{Michael Baggaley}
\author[inst1,inst2]{Ifaz Haider}
\author[inst1,inst2,inst3]{Olivia Bruce}
\author[inst1,inst2,inst4,inst5]{Arash Khassetarash}
\author[inst1,inst2]{W. Brent Edwards}

\affiliation[inst1]{organization={Faculty of Kinesiology, University of Calgary},%Department and Organization
            address line={2500 University Dr. NW}, 
            postcode={T2N 1N4},            
            city={Calgary}, 
            state={Alberta},
            country={Canada}}

\affiliation[inst2]{organization={McCaig Institute for Bone and Joint Health, University of Calgary},%Department and Organization
            address line={3280 Hospital Dr. NW}, 
            postcode={T2N 4Z6},
            city={Calgary}, 
            state={Alberta},
            country={Canada}}
            
\affiliation[inst3]{organization={Department of Radiology, Stanford University},%Department and Organization
            address line={300 Pasteur Dr.}, 
            postcode={94305-5105},
            city={Stanford}, 
            state={California},
            country={USA}}

\affiliation[inst4]{organization={Canadian Sport Institute},%Department and Organization
            address line={151 Canada Olympic Road}, 
            postcode={T3B 6B7},
            city={Calgary}, 
            state={Alberta},
            country={Canada}} 

\affiliation[inst5]{organization={Department of Education and Kinesiology, Vancouver Island University},
            address line={900 Fifth St.}, 
            postcode={V9R 5S5},
            city={Nanaimo}, 
            state={British Columbia},
            country={Canada}}

%TC:ignore
\begin{abstract}
A fatigue-failure process is hypothesized to govern the development of tibial stress fractures, where bone damage is highly dependent on the peak strain magnitude. To date, much of the work examining tibial strains during running has ignored uphill and downhill running despite the prevalence of this terrain. This study examined the sensitivity of tibial strains to changes in running grade and speed using a combined musculoskeletal-finite element modeling routine. Seventeen participants ran on a treadmill at $\pm$10\degree{}, $\pm$5\degree{}, and 0\degree{}; at each grade, participants ran at 3.33 ms\textsuperscript{-1} and at a grade-adjusted speed – 2.50 and 4.17 ms\textsuperscript{-1} for uphill and downhill grades, respectively. Force and motion data were recorded in each grade and speed combination. Muscle and joint contact forces were estimated using inverse-dynamics-based static optimization. These forces were applied to a participant-adjusted finite element model of the tibia. None of the strain variables (50\textsuperscript{th} and 95\textsuperscript{th} percentile strain and strained volume $\geq$4000 $\mu\varepsilon$) differed as a function of running grade; however, all strain variables were sensitive to running speed (\emph{F}$\geq$9.59, \emph{p}$\leq$0.03). In particular, a 1 ms\textsuperscript{-1} increase in speed resulted in a 9$\%$ ($\approx$ 260 $\mu\varepsilon$) and 155$\%$ ($\approx$ 600 mm\textsuperscript{3}) increase in peak strain and strained volume, respectively. Overall, these findings suggest that faster running speeds, but not changes in running grade, may be more deleterious to the tibia. 
\end{abstract}

\begin{highlights}
\item High-magnitude tibial strains that are implicated in stress-fracture development were sensitive to changes in running speed but not grade. 

\item The majority of the variance in measures of tibial strain was accounted for by individual subject variation, reinforcing the importance of inherent musculoskeletal properties in determining the bone strain environment. 
\end{highlights}


\end{frontmatter}

%\begin{linenumbers}

\section{Introduction}
\label{sec:intro}

%%       \citet{<label>} ==> Jones et al. (2015)
%%       \citep{<label>} ==> (Jones et al., 2015)

Tibial stress fractures are thought to result from a fatigue-failure process, wherein repetitive loading and cumulative bouts of activity results in damage accumulation and the progressive weakening of bone \citep{carter1981,Schaffler1989}. Damage accumulation is directly related to the strain magnitude experienced by the bone \citep{carter1981,loundagin2018}. Tibial strains have been well characterized during level ground running using modeling approaches \citep{edwards2009,Meardon2021,rice2019}. Considerably less is known about tibial strains during graded running which is a common component of urban, trail, and mountain running.  

Two studies have examined \emph{in vivo} tibial strains during graded running, both observing lower peak compressive and shear strains in downhill running, compared to uphill running and level jogging \citep{burr1996,milgrom2020}. However, these data were collected on only three participants running at a single grade (20\degree{} in \citep{milgrom2020} and unknown grade in \citep{burr1996}) and an unknown speed. Considering that running biomechanics change with grade and speed \citep{khassetarash2020,vernillo2020}, it is important to quantify tibial strains across a range of running grades and speeds, to characterize the relative risk of graded running. 

The relationship between tibial strain and running speed has been better described, with faster speeds associated with higher strains \citep{Edwards2010,burr1996}. Modeling studies using other tibial loading metrics have also observed increased loading with faster speeds \citep{Meardon2021,rice_speed_2023,baggaley_sensitivity_2022,rice_effect_2023}. However, the relationship between running speed and grade is complex, resulting in different gait adaptations with speed during uphill and downhill running \citep{khassetarash2020}. Further work is needed to understand the interaction between running grade and speed on tibial bone strains.

The complex strain environment of the tibia can be estimated using a combined musculoskeletal-finite element modeling routine, allowing informed decisions about the relative risk of different running conditions.
To date, this approach has not been used to characterize tibial strains during graded running. However, a simplified modeling approach, based on beam theory, demonstrated that downhill running resulted in lower internal tibial bending moments compared to level and uphill running \citep{Baggaley2022,rice_speed_2023}. Bending moments are responsible for the majority of the normal stress (i.e., normal to the cross-section) experienced by the tibia \citep{derrick2016,Baggaley2022}, suggesting that downhill running may be less damaging to the skeletal system. However, beam theory models of long bones may be limited by their lack of complex geometry, for it approximates bones as slender beams of constant cross-sectional area \citep{brassey_finite_2013}. Specifically, \citet{brassey_finite_2013} observed that beam theory models of long bones underestimated the stress due to axial loading and overestimated the stress due to bending compared to a finite element model. Capturing the complex geometry of the tibia is necessary to better estimate the stress-strain response of the bone during running, which warrants the use of the finite element method.

The purpose of this study was to examine the influence of grade and speed on tibial strains during running. In accordance with previous \emph{in vivo} measurements \citep{burr1996,Milgrom2021} and modeling estimates \citep{Baggaley2022,rice_speed_2023}, we hypothesized that downhill running would result in lower tibial strains than both level and uphill running. We also hypothesized that faster running speeds would result in greater tibial strains \citep{Edwards2010,baggaley_sensitivity_2022}.

\section{Methods}

\subsection{Participants}
Seventeen participants [8 female, 9 male (Group average = 27 $\pm$ 8 years, 1.72 $\pm$ 0.08 m, 66.8 $\pm$ 9.9 kg)] were recruited following written informed consent. The study was  approved by the University of Calgary Conjoint Health Research Ethics Board (\#REB14-1117). All participants were injury free the six months prior to participation, were regularly performing physical activity including running, and were comfortable running on a treadmill. Each participant visited the lab on two separate occasions. During the first visit, participants were familiarized to the treadmill grades and footwear used in the experimental conditions (Salomon X-Scream 3d, heel height = 33.7 mm, heel-to-toe insole drop = 12.3 mm). Participant height and mass were measured using a balance and stadiometer. Anthropometric measurements of the lower limbs and pelvis were used to estimate segment masses, moments of inertia, and center of mass locations derived from \citet{Vaughan1999}.

\subsection{Experimental Protocol}
At the second visit, seventeen retro-reflective markers were placed on the participants’ right lower limb to create three-dimensional segmental coordinate systems for the pelvis, thigh, shank, and foot as detailed in \citet{Baggaley2020}. A single, experienced researcher performed all marker placements. 

Participants performed a 5-min warm up at a self-selected speed on an instrumented treadmill (Bertec, Columbus, OH). For the experimental protocol, participants ran at each of five grades (0\degree{}, $\pm$5\degree{}, and $\pm$10\degree{}) in a random order. Data gathered during voluntary pacing experiments in an outdoor setting demonstrate that people tend to slow down when running uphill and speed up when running downhill \citep{mastroianni_voluntary_2000,townshend2010}. To account for changes in self-selected running speed at different grades \citep{townshend2010,mastroianni_voluntary_2000}; a variety of speed and grade combinations were tested. Participants were asked to complete a constant speed condition of 3.33 ms$^{-1}$ at all grades, and also a grade-adjusted speed condition during uphill (2.50 ms$^{-1}$) and downhill running (4.17 ms$^{-1}$). The constant speed condition was used to isolate the effect of grade. Foot-strike pattern was not constrained during the running conditions. After the participant had equilibrated to the target speed, 15 s of data were captured in each condition. Motion capture (Vicon Motion Systems Ltd., Oxford, England) and force platform data were collected concurrently using Vicon Nexus software (Vicon Motion Systems Ltd.) at 200 Hz and 1000 Hz, respectively.

\subsection{Musculoskeletal Model}
Muscle and joint contact forces, for nine stance phases, were calculated using an inverse-dynamics-based static optimization routine in Matlab (Mathworks, Natick, MA), as previously described \citep{baggaley_sensitivity_2022}. Briefly, joint angles were used to drive a musculoskeletal model that included lower extremity bones and 45 muscles scaled to individual segment lengths. The model used muscle definitions as described by \citet{arnold2010}. Moment arms, muscle orientations, and velocity-and-length-adjusted maximal dynamic muscle forces were calculated for each one percent of stance. 
A set of muscle forces were selected that minimized the sum of muscle stresses squared for each frame of data \citep{Crowninshield1981}. Muscle forces were constrained such that the sum of the muscle moments at each joint were equal to the net joint moments about all constrained degrees of freedom. The joint moments used in the optimization were the flexion-extension and abduction-adduction moments at the hip, the flexion-extension moment at the knee and ankle, and the supination-pronation moment at the subtalar joint \citep{baggaley_sensitivity_2022}. Upper and lower bound solutions for muscle forces were set to the velocity-and-length-adjusted maximal dynamic muscle force and zero, respectively. Calculated muscle forces were transformed into the tibial coordinate system and vector summed with the ankle joint reaction force from inverse dynamics analysis to obtain the ankle joint contact force. Joint contact forces were represented in the distal segment coordinate system but were translated and rotated into the tibial coordinate system for finite element modeling.

\subsection{Finite Element Model}
\subsubsection{Participant-Adjusted Finite Element Mesh}
Participant-adjusted finite element meshes were created for each participant using a previously-developed statistical appearance model characterizing tibial-fibular geometry and bone density distribution variation in a sample of young, healthy males and females \citep{bruce_tibial-fibular_2022}. Detailed information about the creation of the statistical appearance model can be found in \citet{bruce_tibial-fibular_2022}. Bone scaling information, accounting for relative changes in both bone size and cortical thickness for each individual, was embedded in the statistical appearance model and characterized within the first principal component (PC1); a strong relationship (\emph{r}$^2$ = 0.8) existed between participant height and the PC1 score \citep{bruce_sex_2023}. Sex-specific average meshes, controlled for scaling, were created by perturbing the statistical appearance model along each principal component, except PC1, by the mean male and mean female scores \citep{bruce_tibial-fibular_2022}. To generate finite element meshes for the new individuals in this study, the relevant sex-specific average mesh (i.e. male or female) was scaled for each participant using the relationship between PC1 score and participant height. This resulted in a unique mesh for each participant that accounted for changes in bone scaling associated with their body size and differences in bone geometry and density due to sex. 

Ten-node tetrahedral elements were used in the finite element mesh, and orthotropic linear-elastic material properties were assigned to each element based on the element average density \citep{Rho1996}. These material property definitions have demonstrated excellent agreement between \emph{ex vivo} cadaveric data and finite element model predictions of fracture strength \citep{gray_experimental_2008,Edwards2013}. Boundary constraints were similar to previous work from our group, with pinned constraints at the knee and ankle and complex proximal tibia-fibula joint constraints \citep{bruce_tibial-fibular_2022,haider2020}. 

Using the statistical appearance model to generate the participant-adjusted finite element meshes provides a low-cost alternative for finite element analysis. However, this approach is best suited for studies using a repeated-measures design, where the accuracy of the absolute strain magnitudes are less important than the relative changes between conditions. Previous work has demonstrated strong agreement (\emph{r}$^2$=0.96) in relative changes of tibial strain during a running bout, between participant-adjusted and participant-specific finite element models despite large absolute errors in strain \citep{khassetarash_tibial_2023}. 

\subsubsection{Loads}
The origin or insertion of seventeen muscles which attach to the tibia and/or fibula plus the patellar ligament were identified by aligning each participant's musculoskeletal model geometry and their finite element mesh using an iterative closest points algorithm and mapping each muscle point to the nearest surface node. A concentrated force was applied at each muscle attachment point. The ankle joint contact force was applied as a concentrated force at the ankle centre of rotation. A residual moment term about the sagittal and transverse plane axes was calculated and applied at the ankle centre of rotation \citep{haider2020}. It was calculated as the difference between the ankle joint moment, from inverse dynamics, and the net moment produced about the ankle by all the muscles that attach to the tibia. The residual moment accounted for other sources of torque acting on the tibia that were not captured in our finite element model, such as the effect of the gastrocnemius. A bi-articular muscle that does not attach to the tibia so its contribution to bone bending is not captured in the finite element model. Further information on the residual moment term can be found in \citet{haider2020}. All loads were applied to the tibia according to their value at peak resultant ankle joint contact force. This instance in stance was chosen as it was expected to result in the highest muscle forces, paralleling joint moment profiles.

\subsection{Dependent Variables}
Finite element models were created for each of the nine steps (using loads at peak ankle contact force) recorded at each grade and speed combination ($\approx$ 1377 models) and were solved in Abaqus (v.2019 Dassault Systèmes Simulia Corp.; Providence, USA). Custom Matlab (Mathworks, Natick, MA) scripts were used to calculate pressure-modified von Mises equivalent strain; a modification of the von Mises strain criterion that has been shown to predict failure in quasi-brittle materials that demonstrate compression-tension strength asymmetry \citep{peerlings_gradient-enhanced_1998,de_vree_comparison_1995,Haider2021}. Analysis was limited to the tibial diaphysis, defined as 20–80\% of the length of the tibia. Only elements containing bone, defined by an element density $\geq$0.5 g/cm\textsuperscript{3}, were used for analysis. This threshold was chosen to exclude strains in the bone marrow and other tissue in the medullary canal. Elements within a 1.0 cm radius of the soleus force application and a 0.5 cm radius of other muscle force application points, including transcortical elements, were removed from the analysis due to artefactually high strains ($>$ 10,000 \(\mu\varepsilon\)). The large force applied at the attachment point for the soleus necessitated a larger radius to remove all elements with artefactually high strains. Over 98\% of the elements containing bone in the tibial diaphysis remained for analysis after artefacts were removed. 

The 95\textsuperscript{th} and 50\textsuperscript{th} percentile pressure-modified von Mises equivalent strain (hereafter referred to as percentile strain) were extracted from the tibial diaphysis for analysis. The 95\textsuperscript{th} percentile strain was used to characterize the peak tibial strain to avoid artefactually high strains due to imaging errors and stress concentrations. the 50\textsuperscript{th} percentile was used to provide more information regarding the strain distribution. We also calculated strained volume, which is defined as the total volume of elements experiencing strain $\geq$ 4000 \(\mu\varepsilon\). Strained volume, with a threshold of 4000 $\mu\varepsilon$, has been shown to be a strong predictor of the fatigue life of bone in uniaxial and biaxial loading \citep{Haider2021}. At the material level, \citet{obrien_behaviour_2005} observed rapid microdamage accumulation and subsequent fracture in cyclically-loaded cortical bone at a stress range of 80 MPa, which would correspond to 4000 $\mu\varepsilon$ for an assumed elastic modulus of 20 GPa; samples loaded at lower stress ranges accumulated damage but did not fracture\citep{obrien_behaviour_2005}. The 50\textsuperscript{th} and 95\textsuperscript{th} percentile strain and strained volume of the tibial diaphysis were the primary variables of interest.

To compare the strains estimated with the finite element models to previous \emph{in vivo} data, planar strains were calculated at a reference location on the medial surface of the tibia at the midpoint between the mid-shaft location and 2 cm distal. This location is comparable to the \emph{in vivo} strain gauge locations used in previous studies \citep{burr1996,milgrom2020}. 
The average three-dimensional strain state of the six elements that surrounded the reference location were transformed into a local planar coordinate system with a unit vector normal to the exterior model surface \citep{edwards_finite_2012}. This created a virtual strain gauge with a coordinate system similar to the strain gauge in \citet{burr1996} and \citet{milgrom2020}. The maximum and minimum principal strain and maximum shear strain were calculated at the virtual strain gauge location. 

\subsection{Statistical Analysis}
For each participant, dependent variables from the nine analyzed stance phases were averaged for each condition. Linear mixed-effects models (LMM) were used to test for the effect of grade and speed on 50\textsuperscript{th} percentile strain, 95\textsuperscript{th} percentile strain, and strained volume for the tibial diaphysis; maximum principal strain,  minimum principal strain, and maximum shear strain for the virtual strain gauge location; and the tri-axial ankle joint contact force. Grade was treated as a categorical factor and speed was treated as a continuous factor (fixed effects); a grade x speed interaction term was also included. Random intercepts were included for each participant (random effect). Residuals of each LLM were examined using Q-Q plots, and the assumption of normality was tested using the Shaprio-Wilk test. If the LMM residuals were not normally distributed, the dependent variable in the LMM was log-transformed and the models were re-fit. The overall coefficient of variation (\emph{r}$^2$) and the median error (\%) between LMM predictions and finite element model measures was calculated to assess model fit for the primary variables of interest - 50\textsuperscript{th} and 95\textsuperscript{th} percentile strain and strained volume.

Statistical significance of the interaction and main effects were determined using a two-way ANOVA ($\alpha$=0.05). In the case of a significant speed x grade interaction, the speed variable in the LMM was re-coded on a 0-1 scale, where 0 represented the 3.33 ms\textsuperscript{-1} condition and 1 represented the grade-adjusted speed condition. To interpret the interaction, pairwise comparisons were performed between grades at each speed and between speeds at each grade. If no interaction was present, the grade x speed interaction term was removed from the LMM. This was needed to characterise the effect of grade as the full model containing the interaction term could not produce an estimated marginal mean for level running since it contained data at only one speed. A significant main effect of grade was interpreted with pairwise comparisons between grade conditions. A Holm-Bonferroni correction was used for all pairwise comparisons \citep{abdi_holms_2010}.
In the case of a significant main effect of speed, the model coefficients ($\beta$ estimate) were examined to determine the effect for every 1 ms\textsuperscript{-1} increase in running speed.  

Statistical analyses were performed in R (R Core Team, 2018) within RStudio (Version 2021.09.1+372, RStudio, PBC, Boston, MA) using \emph{lme4}, \emph{lmerTest}, \emph{effects}, \emph{emmeans}, \emph{tidyverse}, and \emph{rsq} packages \citep{fox_r_2019,wickham_welcome_2019,Bates2014,Kuznetsova2017,Russell2018,zhang_rsq_2022}.

\section{Results}
\begin{figure}
    \centering
    \includegraphics[scale=0.65]{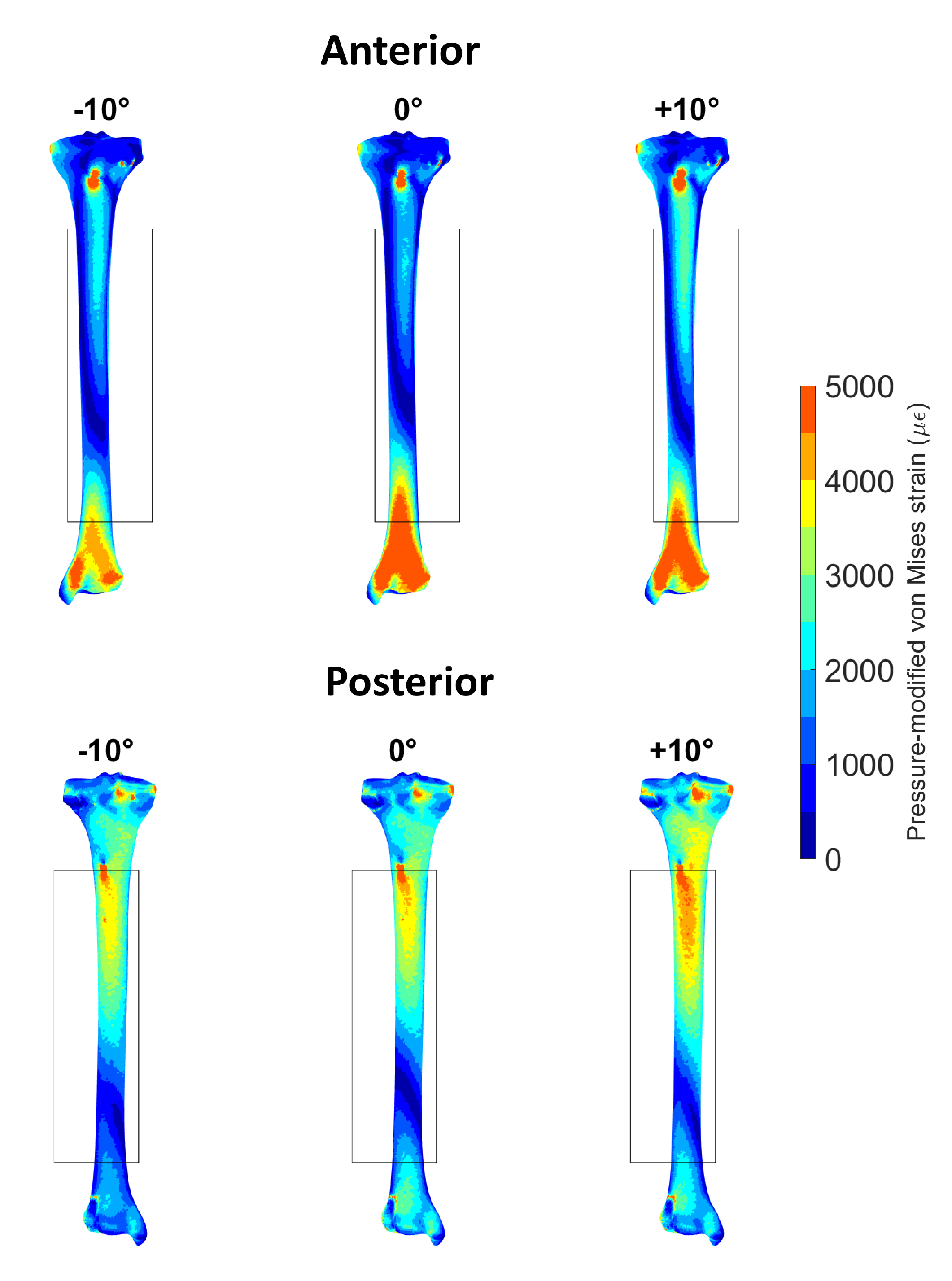}
    \caption{Pressure-modified von Mises equivalent strain distribution of the tibia, from an exemplary participant during $\pm$10\degree{} and 0\degree{} at 3.33 ms\textsuperscript{}. The bounding box indicates the portion of the tibia that was included for the calculation of 50\textsuperscript{th} and 95\textsuperscript{th} percentile strain and strained volume.}
    \label{fig:strain_pmvm}
\end{figure}

\begin{figure}
    \centering
    \includegraphics[scale=0.5]{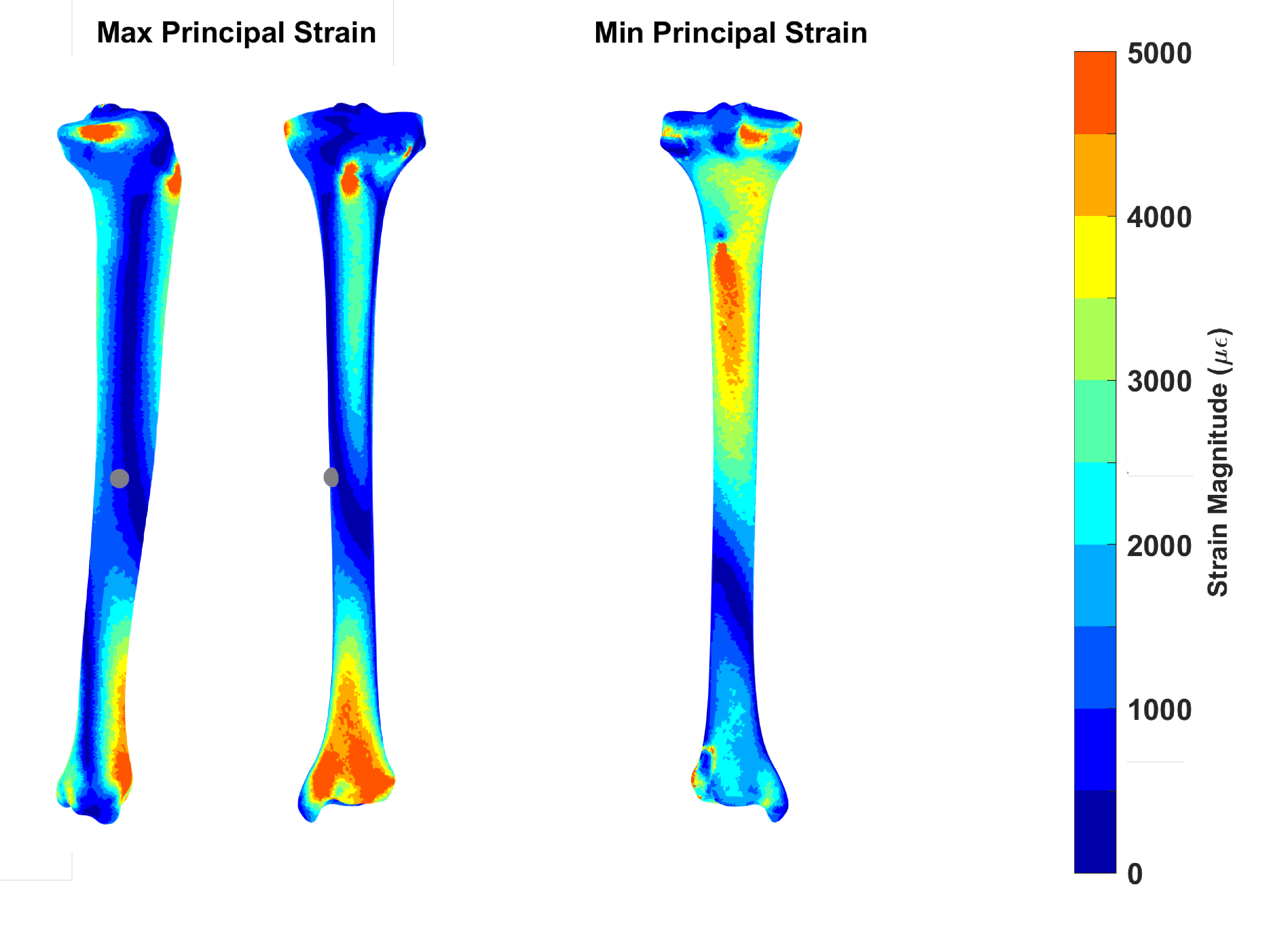}
    \caption{Maximum (left) and minimum (right) principal strain distribution of the tibia from an exemplary participant during 0\degree{} at 3.33 ms\textsuperscript{-1}. A medial and anterior view are presented for the maximum principal strain and a posterior view is provided for the minimum principal strain. Two views are provided for maximum principal strain to indicate the location of the virtual strain gauge (gray sphere), where maximum and minimum principal strain were quantified.}
    \label{fig:strain_dist}
\end{figure}

The pressure-modified von Mises equivalent strain distribution of the full finite element model for an exemplary participant during $\pm$10\degree{} and 0\degree{} at 3.33 ms\textsuperscript{-1} are provided in Figure \ref{fig:strain_pmvm}.
The maximum and minimum principal strain distribution of the full finite element model for an exemplary participant running at 0\degree{} and 3.33 ms\textsuperscript{-1} are provided in Figure \ref{fig:strain_dist}. Descriptive statistics and the results of statistical tests for strain measures are presented in Table \ref{table:strainsTable}, and ankle joint contact force and residual moment data are presented in Table \ref{table:strainsTable2}. Only significant interactions, main effects, and post-hoc comparisons are stated in the results. The $\beta$ estimates and 95\% confidence intervals of the estimates for all dependent variables are presented in the appendix. 

\subsection{Linear mixed-effect model parameters and fit}
Residuals of the linear mixed-effect models violated the assumption of normality for 95\textsuperscript{th} percentile strain, strained volume, and maximum and minimum principal strain; these variables were log-transformed and the models were re-fit. The log-transformed models met the normality criterion. 

Model fit (\emph{r}$^2$) ranged from 0.55 - 0.68 for 50\textsuperscript{th}, 95\textsuperscript{th} percentile strain, and strained volume. In all models, the random effects due to individual variation accounted for the majority of the variance in measures of tibial strain (range = 43 - 64\%). The fixed effects of grade and speed accounted for only 4 - 12\% of the variance in measures of tibial strain. Median error between finite element model measures and LMM predictions were 6.5\% for 50\textsuperscript{th} percentile strain, 4.4\% for 95\textsuperscript{th} percentile strain, and 39\% for strained volume. 

\begin{figure}[h]
    \centering
    \includegraphics[width=\textwidth]{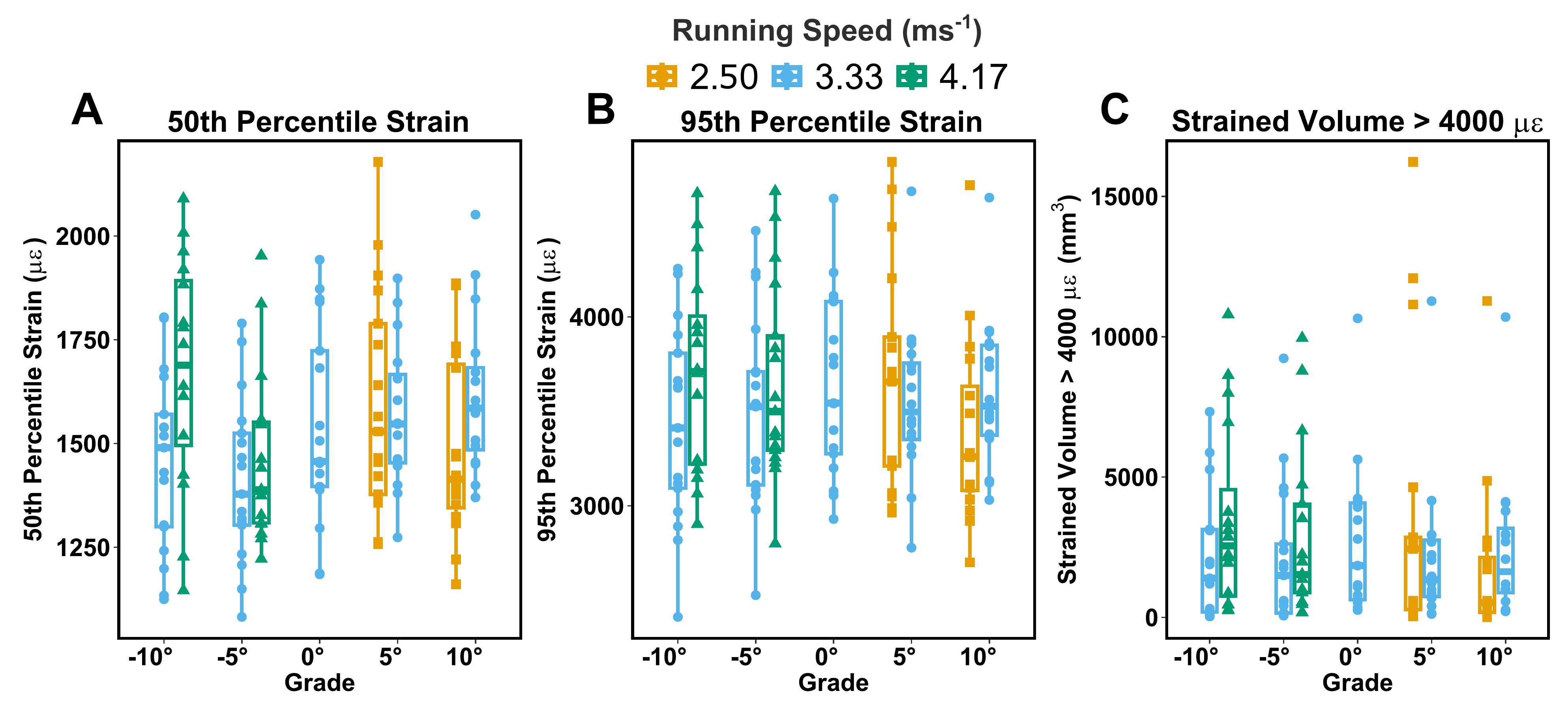}
    \caption[Measures of tibial strain during grade and speed conditions.]{50\textsuperscript{th} (A) and 95\textsuperscript{th} (B) percentile pressure-modified von Mises equivalent strain and strained volume (C) for each grade and speed condition. Box plots are used to display the data with individual participant data (n=17) overlaid.}
    \label{fig:strains.speed.grade}
\end{figure}

\begin{figure}[h]
    \centering
    \includegraphics[width=\textwidth]{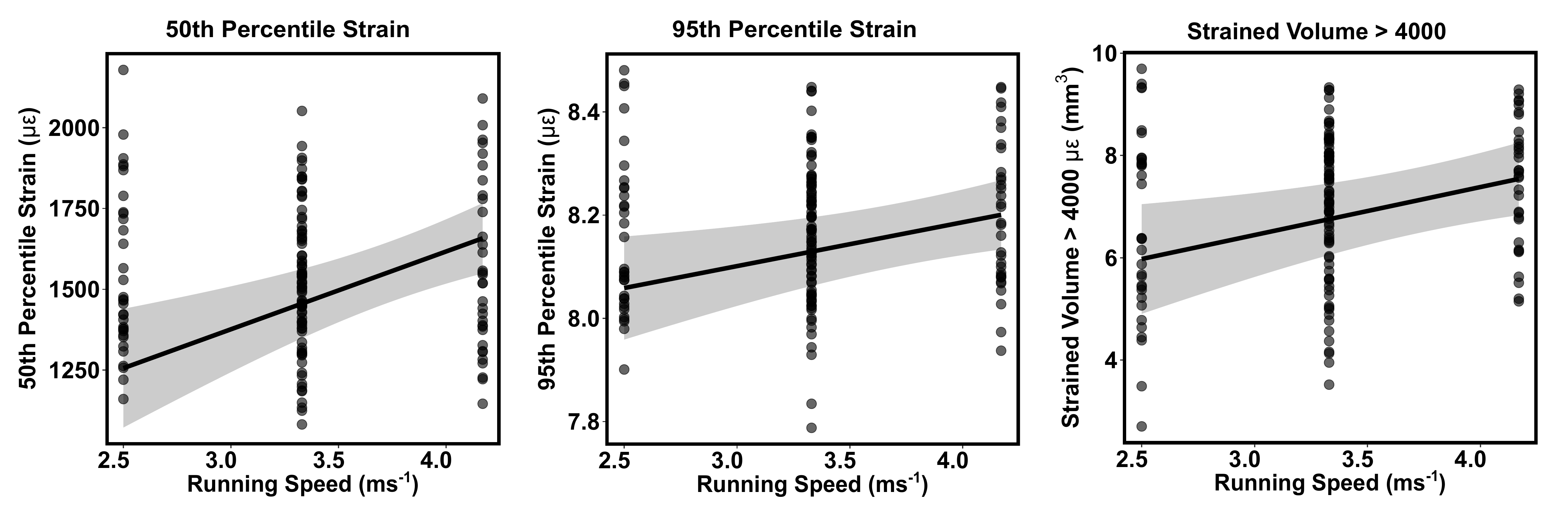}
    \caption[Visualize main effect of speed.]{These plots visualise the slope coefficients ($\beta$ estimate) for the main effect of speed on 50\textsuperscript{th} percentile strain (A), 95\textsuperscript{th} percentile strain, and strained volume $\geq$ 4000 $\mu\varepsilon$. Individual data points for all grade conditions are included to demonstrate that the model predictions matched the experimental data. Please note that the y-axis values for 95\textsuperscript{th} percentile strain and strained volume have been log-transformed (base e; please multiply by 2.178 to get value on the response scale).}
    \label{fig:strains.speed}
\end{figure}

\subsection{Grade x Speed} 
A grade x speed interaction was not observed for any dependent variable (\emph{F}(3)$\leq$2.97, \emph{p}$\geq$0.24).

\subsection{Grade}
A main effect of grade was not observed for any strain variable - tibial diaphysis strains or principal strains at the virtual strain gauge location (\emph{F}(4)$\leq$2.53, \emph{p}$\geq$0.233). 

\paragraph{Ankle Joint Contact Force}
A main effect of grade was observed for axial (\emph{F}(4)=10.66, \emph{p}$<$0.001), anterior-posterior (\emph{F}(4)=94.12, \emph{p}$<$0.001), and medial-lateral ankle joint contact force (\emph{F}(4)=16.69, \emph{p}$<$0.001) (Figure \ref{fig:ACF_grade}). Post-hoc testing revealed that axial ankle joint contact force was lower in both downhill conditions compared to level and +5\degree{} (\emph{t}$\geq$3.49, \emph{p}$\leq$0.002). Anterior-posterior ankle joint contact force was lower in both downhill conditions compared to both uphill conditions (\emph{t}$\geq$12.22, \emph{p}$<$0.001) and level running (\emph{t}$\geq$7.52, \emph{p}$<$0.001), and +10\degree{} running conditions were greater than level running (\emph{t}=3.08, \emph{p}=0.004). Medial-lateral ankle joint contact force was lower in both downhill conditions compared to +5\degree{} (\emph{t}$\geq$5.00, \emph{p}$\leq$0.001) and 0\degree{} (\emph{t}$\geq$4.37, \emph{p}$\leq$0.001). -10\degree{} was also lower than +10\degree{} (\emph{t}=6.53, \emph{p}<0.001), and +5\degree{} was greater than +10\degree{} (\emph{t}=3.15, \emph{p}=0.003).

\subsection{Speed}
\paragraph{Tibial Diaphysis Strains}
A main effect of speed was observed for 50\textsuperscript{th} percentile strain (\emph{F}(1)=10.90 \emph{p}=0.017), 95\textsuperscript{th} percentile strain (\emph{F}(1)=9.59 \emph{p}=0.031) and strained volume (\emph{F}(1)=19.83, \emph{p}$<$0.001). For every 1 ms\textsuperscript{-1} increase in speed, 50\textsuperscript{th} percentile strain increased by 19\% ($\approx$ 240 $\mu\epsilon$), 95\textsuperscript{th} percentile strain increased by 9\% ($\approx$ 260 $\mu\epsilon$) and strained volume increased by 155\% ($\approx$ 600 $\mu\epsilon$).

\paragraph{Virtual Strain Gauge}
A main effect of speed was observed for maximum principal strain (\emph{F}(1)=16.96 \emph{p}=0.001), minimum principal strain (\emph{F}(1)=31.55 \emph{p}$<$0.001), and maximum shear strain (\emph{F}(1)=26.21 \emph{p}$<$0.001). For every 1 ms\textsuperscript{-1} increase in speed, maximum principal strain increased by 60\% ($\approx$ 360 $\mu\epsilon$), minimum principal strain increased by 29\% ($\approx$ 643 $\mu\epsilon$), and maximum shear strain increased by 125\% ($\approx$ 1247 $\mu\epsilon$).

\paragraph{Ankle Joint Contact Force}
A main effect of speed was observed for axial (\emph{F}(1)=17.68, \emph{p}$<$0.001), anterior-posterior (\emph{F}(1)=33.89, \emph{p}$<$0.001), and medial-lateral (\emph{F}(1)=12.26, \emph{p}$<$0.001) ankle joint contact force. For every 1 ms\textsuperscript{-1} increase in speed, axial ankle joint contact force increased by $\approx$ 448 N, anterior-posterior joint contact force decreased by $\approx$ 291 N, and medial-lateral joint contact force increased by $\approx$ 192 N.

\begin{figure}[h]
    \centering
    \includegraphics[width=\textwidth]{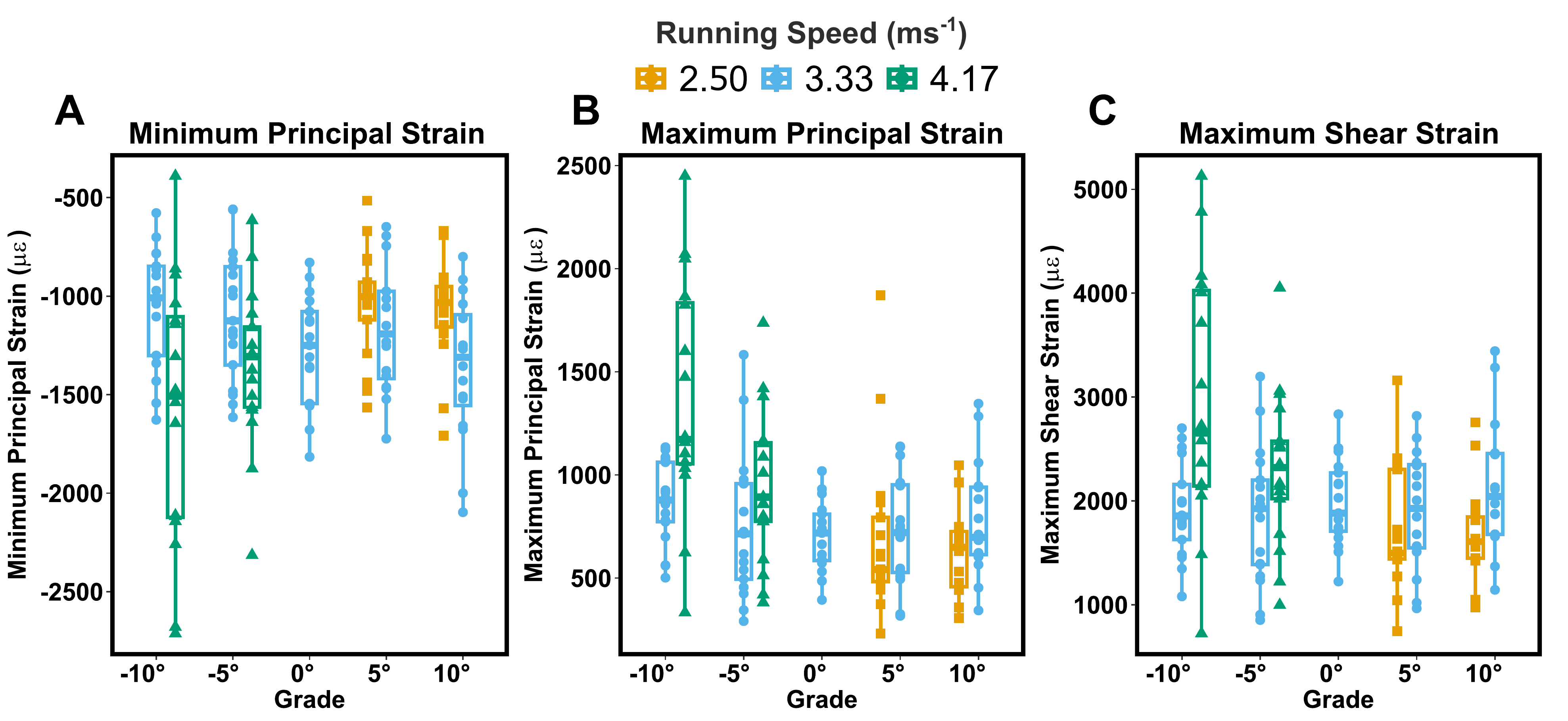}
    \caption{Maximum principal strain (A), minimum principal strain (B), and maximum shear strain (C) for each grade and speed condition. Box plots are used to display the data with individual participant (n=17) data overlaid.}
    \label{fig:Pstrain.speed.grade}
\end{figure}

\section{Discussion}
This study investigated the influence of running grade and speed on tibial strain using a combined musculoskeletal-finite element modeling approach. We hypothesized that downhill running would result in lower strains than uphill and level running. Contrary to our hypothesis, we observed no effect of grade on tibial diaphysis strains nor principal strains at the virtual strain gauge location. In contrast, all measures of tibial strain were sensitive to running speed. These data suggest that faster running speeds, but not changes in grade, may be associated with an increased risk of tibial stress fracture. 

\begin{figure}
    \centering
    \includegraphics[width=\textwidth]{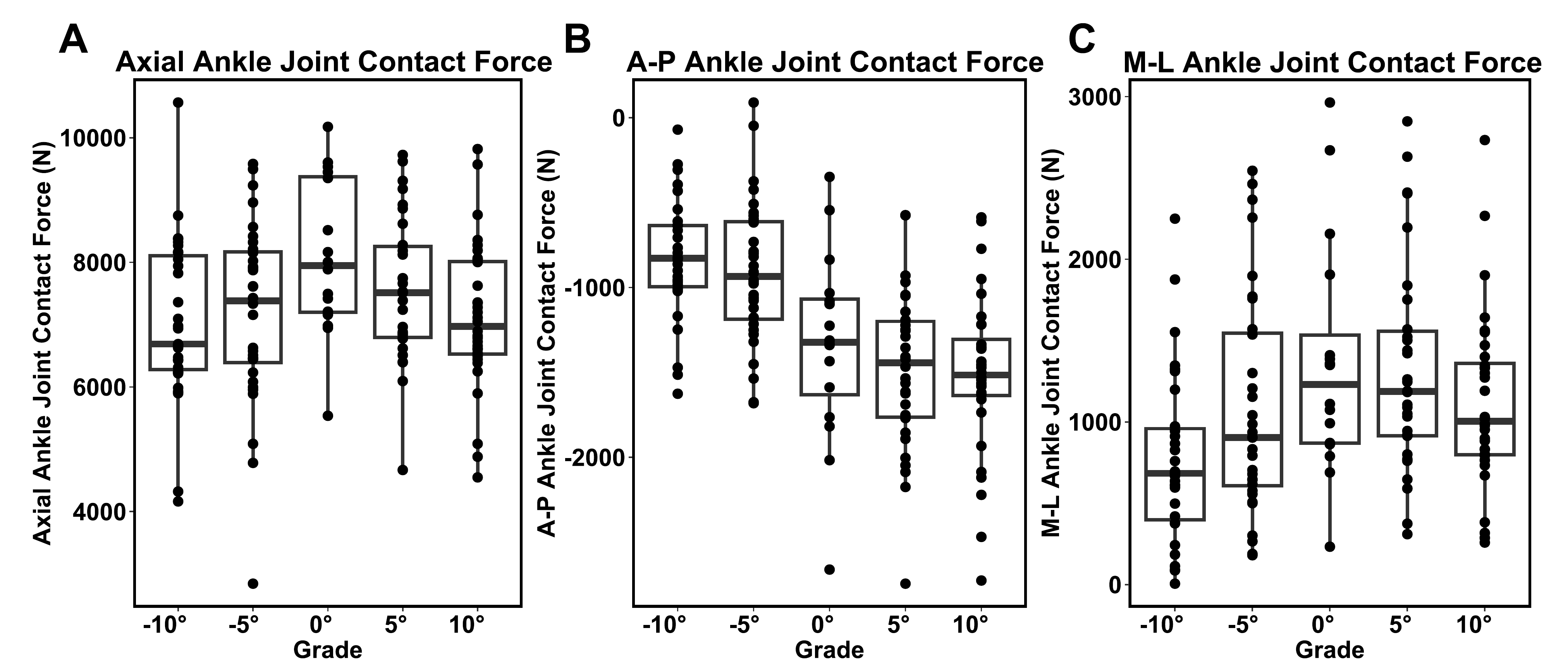}
    \caption[Peak ankle joint contact force in all grade conditions]{Peak ankle joint contact force in the axial (A), anterior-posterior (B), and medial-lateral (C) direction in all grade conditions collapsed across speed. Box plots are used to display the data with individual participant data (n=17) overlaid.}
    \label{fig:ACF_grade}
\end{figure}

In the development of stress-fractures, high-magnitude strains are thought to be responsible for the fatigue-failure process that ultimately results in injury \citep{burr1985}. Since no effect of grade was observed for 95\textsuperscript{th} percentile strain or strained volume (Figure \ref{fig:strains.speed.grade}B and C), this suggests that graded running may have little effect on the risk of tibial stress fracture. The results of this study contradict the conclusions drawn from simpler modeling approaches based on beam theory \citep{Baggaley2022,rice_speed_2023}. However, these models were not able to capture the complex stress-strain response of the whole tibia, as they did not incorporate the three-dimensional geometry of the tibia \citep{bruce_sex_2023}. By incorporating the geometry of the tibia, it was observed that the tibial strain distribution was relatively insensitive to the large alterations in ankle joint contact force with running grade (Figure \ref{fig:ACF_grade}). While previous studies have suggested that measures of tibial load, based on joint contact forces, are better representatives of the internal loading environment than ground-reaction force parameters \citep{walker_tibial_2022,matijevich2019}; the results of this study demonstrate that erroneous conclusions can still be drawn when looking at joint contact forces alone. Instead, the variation in tibial strain was largely (43 - 64$\%$) due to between-subject variation (i.e., bone geometry and running biomechanics).

\begin{figure}[h]
    \centering
    \includegraphics[width=\textwidth]{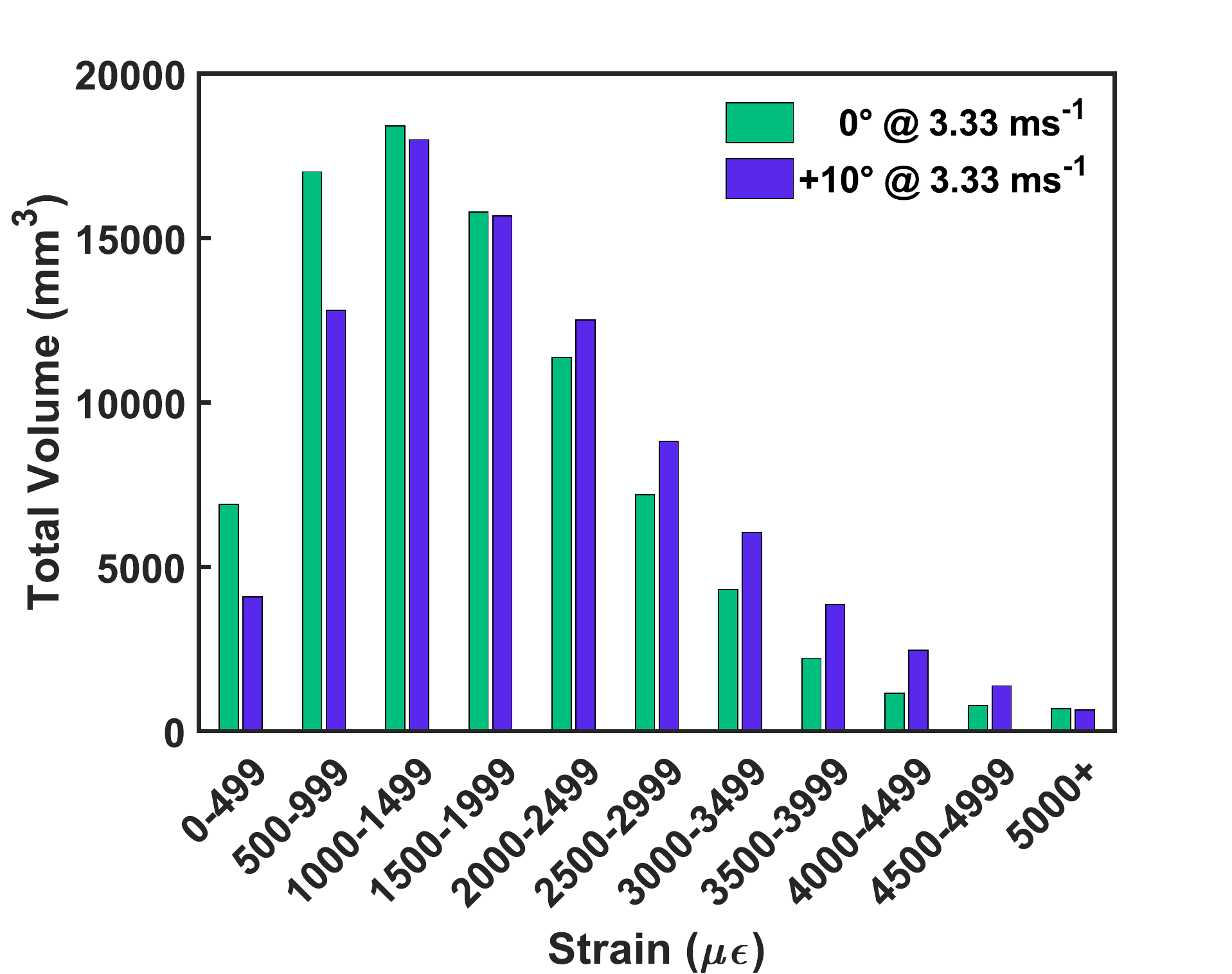}
    \caption{The pressure-modified von Mises equivalent strain distribution of the tibial diaphysis in 0\degree{} and +10\degree{} conditions for the participant data in Figure \ref{fig:strain_pmvm}. The strain distribution was divided into 500 \(\mu\epsilon\) bins, and the total volume of elements experiencing strains between the upper and lower bounds of each bin were quantified.}
    \label{fig:StrainDist_s13}
\end{figure}

Tibial strains were sensitive to speed at all grades of running; faster running speeds were associated with larger high- and low-magnitude strains (\ref{fig:strains.speed}). To account for natural fluctuations in speed when running on graded terrain \citep{townshend2010,mastroianni_voluntary_2000}, a grade-adjusted speed condition was included, with a faster running speed used for downhill conditions (4.17 ms\textsuperscript{-1}) and a slower speed used for uphill conditions (2.50 ms\textsuperscript{-1}). By including these data, the results of the study demonstrate that it is the changes in running speed, made when traversing graded terrain, that are likely to determine the damage potential to the tibia. While changes in running grade are associated with variable changes to running biomechanics \citep{khassetarash2020,vernillo2020}, changes with speed are more uniform and can help us understand why tibial strains increased with running speed. Increasing running speed is typically associated with an increased stride length \citep{bailey_is_2017,orendurff_little_2018}, increased stride frequency \citep{vernillo2020,bailey_is_2017}, and a concomitant increase in knee and ankle joint torques \citep{orendurff_little_2018,Schache2011,khassetarash2020}. Increased knee and ankle joint torques are achieved by producing larger forces from the muscles surrounding the knee and ankle joints; this will necessarily increase the forces applied to the tibia, resulting in the higher tibial strains observed in this study. For a more complete description of the gait alterations made during the grade and speed conditions used in this study, readers are referred to \citet{khassetarash2020}, as the same experimental data was used for both manuscripts.         

The conclusions drawn herein are based on the hypothesis that stress fractures are a fatigue-failure phenomenon where the fatigue process is highly dependent on strain magnitude \citep{Haider2021,zioupos2001,carter1981}. Strained volume has been shown to be a better predictor of the fatigue-life of bone in bi-axial loading than peak strain \citep{Haider2021}; thus, strained volume may be most informative to infer the risk of tibial stress fracture. The results of this study suggest that graded running may have little effect on the risk of tibial stress fracture. In contrast, faster running speeds are likely to increase the risk of tibial stress fracture given the large (155\%, $\approx$ 600 mm\textsuperscript{3}) increase in strained volume associated with every 1 ms\textsuperscript{-1} increase in speed. This is corroborated by \emph{in vivo} and \emph{in silico} studies demonstrating that faster running speeds are associated with greater tibial strains and increased risk of tibial stress fracture in humans \citep{Edwards2010,burr1996} and other mammals \citep{Biewener1986}. 

Fatigue-failure results from damage accumulation in the bony matrix due to repetitive loading over many bouts of activity. The interpretation of the results are based on tibial strains from a single step; however, when running uphill individuals tend to adopt a higher step frequency \citep{vernillo2017}, and this would result in more loading cycles applied to the tibia when running a fixed distance. Thus, uphill running may result in greater cumulative damage to the tibia, given that the 95\textsuperscript{th} percentile strain and the strained volume $\geq$ 4000 \(\mu\varepsilon\) were similar between grades. In future studies, theoretical models of damage accumulation in bone \citep{Taylor2004} could be used to incorporate cumulative effects of running to predict the probability of stress fracture in running programs with different speed and grade combinations.  

\subsection{Limitations}
Musculoskeletal modeling is based on a large number of simplifying assumptions, and it is difficult to know the true impact the assumptions have on our estimates of internal musculoskeletal forces. Recent work has demonstrated the importance of using physiologically realistic boundary conditions in finite element models \citep{haider2020}; however, much work is left to be done to determine the most appropriate way to model the loads and boundary conditions to approximate the \emph{in vivo} environment of the tibia. 

To compare our results with \emph{in vivo} bone strains during graded running \citep{burr1996,milgrom2020}, planar strains (maximum and minimum principal strain and shear strain) were quantified on the medial surface of the tibia at a location corresponding to the strain gauge location used by \citet{burr1996} and \citet{milgrom2020}. The minimum principal strain magnitudes from the finite element model (-1037 to -1350 \(\mu\varepsilon\)) were similar to those reported by \citet{milgrom2020} (-798 to -1528 \(\mu\varepsilon\)) while the maximum principal strains were approximately half the magnitude (541 to 1171 \(\mu\varepsilon\)) reported by \citet{milgrom2020} (1480 to 1691 \(\mu\varepsilon\)) but more closely matched those reported by \citet{burr1996} (625 to 707 \(\mu\varepsilon\)). The \emph{in vivo} strain gauge data found that minimum principal strain and maximum shear strain were lower during downhill running than level and uphill running \citep{burr1996,milgrom2020}. In partial agreement, minimum principal strain was lower during downhill running compared to shallow uphill running at a constant speed in the present study. The similarities between the three data sets provides confidence in the modeling routine to produce strains that are similar to those measured \emph{in vivo}. However, disparate results between planar surface strains and measures of the whole-bone strain distribution demonstrate the importance of quantifying the tibial strain distribution to estimate the relative risk of stress fracture in different running conditions. It is important to note, that the conclusions of this study have not been validated against prospective injury data, and thereby represent our prediction of the relative risk of tibial stress fracture as a function of running grade and speed.

The present study is also limited by the experimental protocol employed. Graded running on a treadmill may not reflect how individuals engage with graded terrain when running in an outdoor environment. While a range of speeds was used in the grade conditions to capture differences in running speed observed in outdoor environments, it is likely that the chosen speeds do not reflect how some participants would freely choose to engage with graded terrain. A self-selected speed condition would be an informative addition to determine how individual running strategies influence tibial strains. A non-linear speed vs grade response has also been observed in outdoor running during voluntarily paced activities \citep{townshend2010,mastroianni_voluntary_2000}. Short uphill sections can be traversed at speeds which would be unsustainable during longer uphill sections due to the slow VO\textsubscript{2} response to changes in exercise intensity \citep{Minetti2020}. In contrast, on short downhill sections, individuals could take advantage of their potential energy and allow themselves to continuously accelerate downhill. The pacing strategy chosen likely has to do with the length of the hill \citep{minetti2002,mastroianni_voluntary_2000} and the experience of the runner. Nonetheless, we hypothesize that the observed speed relationships would hold true for individuals running faster or slower than in this study, producing higher and lower strains, respectively.

Finally, the treadmill presents a relatively constrained environment compared to running outdoors, and individuals may adopt a more conservative gait strategy in situations with greater uncertainty. Future studies could employ wearable technology to determine whether graded running biomechanics are adequately captured in a treadmill environment. 

\section{Conclusion}
Tibial strains were sensitive to changes in running speed but not grade, suggesting that on a per-step basis, running faster may increase the risk of tibial stress fracture, unlike running on graded terrain. In the broader context of understanding stress fracture risk between individuals, it is important to highlight that tibial strains were predominantly influenced by individual participant differences (i.e., bone geometry and running biomechanics), rather than the effect of running grade or speed.

\section{Conflict of interest statement}
The authors have no conflicts of interest to declare. 

\section{Funding}
This work was funded in part by the Natural Sciences and Engineering Research Council of Canada (RGPIN 01029-2015, 02404-2021).

%\end{linenumbers}

% \appendix

\section{Appendix}

\subsection{Linear Mixed Model Outputs}

\begingroup\small
\setlength{\LTcapwidth}{6in}
\begin{longtable}{ccc}
\caption{Beta estimates and lower and upper limits for 95 \% confidence intervals for linear mixed model parameters. The value of the intercept corresponds to the value of the variable in the -10 \degree{} condition. Estimates and confidence intervals are presented as the log-transformation of the variable (as used in the mixed model) for 95\textsuperscript{th} percentile strain, strained volume, and maximum principal strain. Two linear mixed models were needed to characterise the main effects for anterior-posterior ankle joint contact force (A-P AJCF) as a model containing both grade and speed incorrectly predicted an inverse relationship with speed that did not match the experimental data.} \label{table:appendix}\\
    \hline
     & Estimate ($\beta$) & 95\% Confidence Intervals \\
     \hline
     \endfirsthead
     \hline
     & Estimate ($\beta$) & 95\% Confidence Intervals \\
     \hline
     \endhead
    && Continued on next page \\
    \hline
     \endfoot
     \hline
     \endlastfoot
         \textbf{50\textsuperscript{th} Percentile Strain ($\mu\varepsilon$)} &  & \\
         Intercept&654.69 &183.86, 1125.34 \\
         -5 \degree{}&549.15&-103.99, 1202.51 \\
         0 \degree{}&85.71&-16.42, 187.85 \\
         5 \degree{}&587.09&52.97, 1121.37 \\
         10 \degree{}&991.21&459.63, 1523.13 \\
         Speed&240.56 &116.90, 364.28 \\
         -5 \degree{} x Speed&-178.33  &-351.80, -4.91 \\
         5 \degree{} x Speed&-84.46  &-261.66, 92.67 \\
         10 \degree{} x Speed&-270.365  &-446.48, -94.38 \\
         \textbf{95\textsuperscript{th} Percentile Strain ($\mu\varepsilon$)} &  & \\
         Intercept&7.85 &7.61, 8.08 \\
         -5 \degree{}&0.12&-0.21, 0.45 \\
         0 \degree{}&0.06&0.011, 0.11 \\
         5 \degree{}&0.14&-0.13, 0.41 \\
         10 \degree{}&0.37&0.11, 0.64 \\
         Speed&0.08 &0.02, 0.15 \\ 
         -5 \degree{} x Speed&-0.03  &-0.12, 0.06 \\
         5 \degree{} x Speed&0.00  &-0.09, 0.08 \\
         10 \degree{} x Speed&-0.10  &-0.19, -0.02 \\
         \textbf{Strained Volume $\geq$ 4000 $\mu\varepsilon$ (mm\textsuperscript{3})} &  & \\
         Intercept&3.63 &1.05, 6.21 \\
         -5 \degree{}&0.49&-3.06, 4.04 \\
         0 \degree{}&0.61&0.06, 1.17 \\
         5 \degree{}&0.69&-2.21, 3.60 \\
         10 \degree{}&2.93&0.07, 5.85 \\
         Speed&0.94 &0.26, 1.61 \\
         -5 \degree{} x Speed&-0.16  &-1.10, 0.78 \\
         5 \degree{} x Speed&0.27  &-0.69, 1.23 \\
         10 \degree{} x Speed&-0.72  &-1.67, -0.24 \\         
         \textbf{Maximum Principal Strain ($\mu\varepsilon$)} &  & \\
         Intercept&5.17 &4.16, 6.19 \\
         -5 \degree{}&0.35&-1.07, 1.78 \\
         0 \degree{}&-0.211&-0.43, 0.01 \\
         5 \degree{}&0.78&-0.38, 1.95 \\
         10 \degree{}&1.03&-0.13, 2.19 \\
         Speed&0.47 &0.20, 0.74 \\
         -5 \degree{} x Speed&-0.17  &-0.54, 0.21 \\
         5 \degree{} x Speed&-0.27  &-0.61, 0.06 \\
         10 \degree{} x Speed&-0.37  &-0.71, -0.03 \\
         \textbf{Minimum Principal Strain ($\mu\varepsilon$)} &  & \\
         Intercept&1092.26 &143.83, 2040.68 \\
         -5 \degree{}&-1192.61&-2518.12, 132.93 \\
         0 \degree{}&-226.41&-433.68, -19.14 \\
         5 \degree{}&-1624.73&-2708.42, -540.56 \\
         10 \degree{}&-1825.98&-2905.33, -747.52 \\
         Speed&-643.41 &-894.35, -392.47 \\
         -5 \degree{} x Speed&343.73  &-8.21, 695.66 \\
         5 \degree{} x Speed&386.07  &71.87, 700.27 \\
         10 \degree{} x Speed&505.54  &192.94, 818.55 \\
         \textbf{Maximum Shear Strain ($\mu\varepsilon$)} &  & \\
         Intercept&-2242.03 &-4036.04, -488.23 \\
         -5 \degree{}&2322.54 &-189.35, 4834.66 \\
         0 \degree{}&64.56 &-328.26, 457.38 \\
         5 \degree{}&3087.52 &1033.20, 5141.32 \\
         10 \degree{}&3589.50 &1545.87, 5634.84 \\
         Speed&1247.23 &771.74, 1722.78 \\
         -5 \degree{} x Speed&-714.29  &-1381.27, -47.39 \\
         5 \degree{} x Speed&-849.20  &-1444.65, -253.82 \\
         10 \degree{} x Speed&-1073.74  &-1666.81, -481.40 \\
         \textbf{Axial AJCF (N)} &  & \\
         Intercept&4807.27 &3622.92, 5991.63 \\
         -5 \degree{}&155.58 &-169.74, 480.75 \\
         0 \degree{}&1313.17 &900.52, 1725.70 \\
         5 \degree{}&880.70 &479.14, 1283.06 \\
         10 \degree{}&584.43 &188.02, 980.76 \\
         Speed&591.34 &315.02, 867.75 \\
         \textbf{A-P AJCF (N) - Model with grade only} &  & \\
         Intercept&-826.83 &-1044.53, -609.08 \\
         -5 \degree{}&-86.80 &-177.47, 3.89 \\
         0 \degree{}&-512.82 &-623.54, -402.08 \\
         5 \degree{}&-663.86 &-756.42, -571.25 \\
         10 \degree{}&-687.40 &-778.07, -596.71 \\
         Speed&-43.90 &-182.99, 95.17 \\
         \textbf{A-P AJCF (N) - Model with speed only} &  & \\
         Intercept&-2172.09 &-2556.50, -1787.65 \\
         Speed&291.28 &192.82, 389.68 \\         
         \textbf{M-L Ankle Joint Contact Force (N)} &  & \\
         Intercept&29.67 &-465.33, 524.71 \\
         -5 \degree{}&327.51 &199.98, 454.99 \\
         0 \degree{}&691.54 &529.78, 853.26 \\
         5 \degree{}&734.35 &576.77, 891.91 \\
         10 \degree{}&524.05 &368.65, 679.42 \\
         Speed&192.99 &84.67, 301.35 \\
    \end{longtable}

\bibliographystyle{elsarticle-harv} 
\bibliography{references}





% \end{thebibliography}
\begin{landscape}

\begin{table}
 \renewcommand{\arraystretch}{1.25}
\tabcolsep=0.04cm
\scriptsize
\caption{50\textsuperscript{th} and 95\textsuperscript{th} percentile pressure-modified von Mises equivalent strain, strained volume, maximum and minimum principal strain, and maximum shear strain for each grade and speed condition. Mean and (standard deviation) are presented for 50\textsuperscript{th} percentile pressure-modified von Mises equivalent strain, minimum principal strain, and shear strain, and median and (interquartile range) are presented for 95\textsuperscript{th} percentile pressure-modified von Mises equivalent strain, strained volume, and maximum principal strain. Test statistics are provided for the interaction and main effects of grade and speed.}
\label{table:strainsTable}
\noindent
\begin{tabularx}{1.3\textwidth}{m{11em}cccccccccc}
 \hline \hline
&\multicolumn{2}{c}{-10\degree{}}& \multicolumn{2}{c}{-5\degree{}}& 0\degree{} &\multicolumn{2}{c}{+5\degree{}}& \multicolumn{2}{c}{+10\degree{}}& \\

 &3.33 ms\textsuperscript{-1}&4.17 ms\textsuperscript{-1}&3.33 ms\textsuperscript{-1} &4.17 ms\textsuperscript{-1}&3.33 ms\textsuperscript{-1}&2.50 ms\textsuperscript{-1} &3.33 ms\textsuperscript{-1}&2.50 ms\textsuperscript{-1}&3.33ms\textsuperscript{-1}& Test Statistic\\
 \hline
 &&&&&&&&&&Grade x Speed: (\emph{F}(3)=3.23, \emph{p}=0.24)\\
 \multirow{2}{20em}{50\textsuperscript{th} percentile strain (\(\mu\varepsilon\))} &1456&1666&1411&1463&1542&1487&1617&1598&1566&Grade: (\emph{F}(4)=2.87, \emph{p}=0.23) \\
 &(215)&(279)&(200)&(203)&(237)&(224)&(189)&(270)&(173)&Speed: (\emph{F}(1)=10.90, \emph{p}=0.017)\\
 &&&&&&&&&&Grade x Speed: (\emph{F}(3)=2.22, \emph{p}=0.37)\\
 \multirow{2}{20em}{95\textsuperscript{th} percentile strain (\(\mu\varepsilon\))} &3413&3711&3525&3499&3544&3261&3528&3658&3497&Grade: (\emph{F}(4)=2.98, \emph{p}=0.23)\\
 &(716)&(783)&(600)&(605)&(806)&(552)&(475)&(683)&(406)&Speed: (\emph{F}(1)=9.59, \emph{p}=0.03)\\
 &&&&&&&&&&Grade x Speed:(\emph{F}(3)=1.36, \emph{p}=0.45)\\
  \multirow{2}{20em}{Strained Volume (mm\textsuperscript{3})}&1393 &2546&1494&1529 &1844 &527&1629 &2444&1343 &Grade: (\emph{F}(4)=2.32, \emph{p}=0.35)\\
 &(2943)&(3794)&(2462)&(3152)&(3440)&(1963)&(2298)&(2570)&(2011)&Speed: (\emph{F}(1)=19.83, \emph{p}$<$0.001)\\
 &&&&&&&&&&Grade x Speed:(\emph{F}(3)=0.94, \emph{p}=0.42)\\
 \multirow{2}{20em}{Max principal strain (\(\mu\varepsilon\))}
 &878&1172&714&893&718&649&697&542&720&Grade: \emph{F}(4)=2.26, \emph{p}=0.31)\\
 &(289)&(780)&(463)&(382)&(225)&(268)&(327)&(311)&(424)&Speed: \emph{F}(1)=16.96 ,\emph{p}=0.001\\
 &&&&&&&&&&Grade x Speed:(\emph{F}(3)=2.18, \emph{p}=0.25)\\
 \multirow{2}{20em}{Min principal strain (\(\mu\varepsilon\))}
 &-1050&-1590&-1098&-1350&-1277&-1075&-1366&-1037&-1169&Grade: \emph{F}(4)=2.06, \emph{p}=0.31)\\
 &(303)&(685)&(333)&(398)&(282)&(270)&(370)&(278)&(317)&Speed: \emph{F}(1)=31.55 ,\emph{p}$<$0.001\\
 &&&&&&&&&&Grade x Speed:(\emph{F}(3)=2.96, \emph{p}=0.27)\\
 \multirow{2}{20em}{Max shear strain (\(\mu\varepsilon\))}
 &1911&2961&1855&2303&1976&1695&2145&1729&1899&Grade: \emph{F}(4)=2.44, \emph{p}=0.34)\\
 &(460)&(1238)&(656)&(738)&(422)&(459)&(636)&(600)&(565)&Speed: \emph{F}(1)=26.21 ,\emph{p}$<$0.001\\
\hline \hline
\end{tabularx}
\end{table}

\begin{table}
 \renewcommand{\arraystretch}{1.25}
\tabcolsep=0.04cm
\scriptsize
\caption[Ankle joint contact force  during graded running.]{Ankle joint contact force (AJCF) magnitude in  axial, anterior-posterior (A-P), and medial-lateral (M-L) directions for each grade and speed condition. Median and (interquartile range) are presented for Axial and M-L AJCF. Test statistics are provided for the interaction and main effects of grade and speed.}
\label{table:strainsTable2}
\noindent
\begin{tabularx}{1.25\textwidth}{m{8em}cccccccccc}
 \hline \hline
&\multicolumn{2}{c}{-10\degree{}}& \multicolumn{2}{c}{-5\degree{}}& 0\degree{} &\multicolumn{2}{c}{+5\degree{}}& \multicolumn{2}{c}{+10\degree{}}& \\

 &3.33 ms\textsuperscript{-1}&4.17 ms\textsuperscript{-1}&3.33 ms\textsuperscript{-1} &4.17 ms\textsuperscript{-1}&3.33 ms\textsuperscript{-1}&2.50 ms\textsuperscript{-1} &3.33 ms\textsuperscript{-1}&2.50 ms\textsuperscript{-1}&3.33ms\textsuperscript{-1}& Test Statistic\\
 \hline
 &&&&&&&&&&Grade x Speed: (\emph{F}(3)=1.66, \emph{p}=0.18)\\
 \multirow{2}{20em}{Axial AJCF (N)} &6550&6940&6982&7520&7946&6856&7743&6774&7195&Grade: (\emph{F}(4)=10.66, \emph{p}$<$0.001) \\
 &(1740)&(1666)&(2070)&(1921)&(2179)&(1372)&(1671)&(926)&(1672)&Speed: (\emph{F}(1)=17.68, \emph{p}$<$0.001)\\
 &&&&&&&&&&Grade x Speed: (\emph{F}(3)=1.17, \emph{p}=0.33)\\
 \multirow{2}{20em}{A-P AJCF (N)} &-809&-858&-833&-995&-1340&-1370&-1596&-1442&-1586&Grade: (\emph{F}(4)=94.12 \emph{p}$<$0.001)\\
 &(315)&(400)&(435)&(402)&(565)&(406)&(530)&(423)&(530)&Speed: (\emph{F}(1)=33.89, \emph{p}$<$0.001)\\
 &&&&&&&&&&Grade x Speed:(\emph{F}(3)=1.56, \emph{p}=0.20)\\
  \multirow{2}{20em}{M-L AJCF (N)} &652&695&869&972&1230&1033&1189&891 &1111 &Grade: (\emph{F}(4)=16.69, \emph{p}$<$0.001)\\
 &(485)&(564)&(725)&(955)&(665)&(686)&(726)&(604)&(446)&Speed: (\emph{F}(1)=12.26, \emph{p}=0.14)\\
\hline \hline
\end{tabularx}
\end{table}

\end{landscape}

\end{document}